\documentclass[prl,showpacs,twocolumn,superscriptaddress]{revtex4}

\usepackage{graphicx}
\usepackage{amsmath}
\usepackage{amsfonts}
\usepackage{amssymb}
\usepackage{bm}
\usepackage{psfrag}

\newcommand{\dI}[1]{\dot{I}(#1)}
\newcommand{\lrangle}[1]{\langle #1 \rangle}
\newcommand{\LRangle}[1]{\left\langle #1 \right\rangle}

\begin{document}

\title{Detection of finite frequency current moments with a dissipative resonant circuit                                                   }

\author{A. Zazunov}
\affiliation{Centre de Physique Th\'eorique, Universit\'e de la M\'editerran\'ee, Case 907, 13288 Marseille, France}
\affiliation{LPMMC, Av. des Martyrs, 38042 Grenoble, France}

\author{M. Creux}
\affiliation{Centre de Physique Th\'eorique, Universit\'e de la M\'editerran\'ee, Case 907, 13288 Marseille, France}

\author{E. Paladino}
\affiliation{MATIS CNR-INFM, 
and
D.M.F.C.I., Universit\'a di Catania, 95125 Catania, Italy}
\affiliation{Centre de Physique Th\'eorique, Universit\'e de la M\'editerran\'ee, Case 907, 13288 Marseille, France}

\author{A. Cr\'epieux}
\affiliation{Centre de Physique Th\'eorique, Universit\'e de la M\'editerran\'ee, Case 907, 13288 Marseille, France}

\author{T. Martin}
\affiliation{Centre de Physique Th\'eorique, Universit\'e de la M\'editerran\'ee, Case 907, 13288 Marseille, France}

\date{\today}

\begin{abstract}
We consider the measurement of higher current moments with a dissipative resonant circuit, which is coupled inductively 
to a mesoscopic device in the coherent regime. Information about the higher current moments is coded in the histograms
of the charge on the capacitor plates of the resonant circuit. Dissipation is included via the Caldeira-Leggett model, 
and it is essential to include it in order for the charge fluctuations (or the measured noise) to remain finite.
We identify which combination of current correlators enter the measurement of the third moment. The latter
remains stable for zero damping. Results are illustrated briefly for a quantum point contact.    
\end{abstract}


\maketitle

The knowledge of all current moments, at arbitrary frequencies,
allows to characterize completely the statistics of 
electron transfer in mesoscopic devices. 
The lowest current moments have recently been measured 
experimentally for a few specific systems 
\cite{reulet,reznikov,fujisawa}.
Zero frequency noise measurements have provided 
valuable diagnosis for transport in the past, yet
current moments at high frequencies
are difficult to measure, and typically require an on-chip 
measuring apparatus\cite{aguado_kouwenhoven,deblock_science,onac,lesovik_loosen}. 
Finite frequency noise contains information 
which is not apparent at zero frequency, when characterizing
excitations in carbon nanotubes \cite{trauzettel_lebedev}. 
Here, we present a 
scheme for the measurement of the noise and third moment at 
high frequencies, using a resonant circuit. 
A central issue deals with the 
electromagnetic environment on such measurements, 
which has been discussed in the past in different contexts \cite{beenaker,other}. 

On-chip noise measuring 
proposals are either based on capacitive coupling, 
on inductive coupling, or both\cite{gabelli}.
Any measurement involves the filtering 
of frequencies by the detection circuit, with an appropriate 
bandwidth: this justifies the choice of a generic 
resonant circuit.
A dissipationless LC circuit 
was proposed\cite{lesovik_loosen} to measure high frequency noise. 
The measured noise (the squared charge fluctuations on the capacitor) 
is then a combination of 
the unsymmetrized current correlators.
The charge fluctuations are inversely proportional to
the adiabatic switching parameter 
used for the coupling. 
This parameter has thus to be interpreted as a 
line width which should be computed from first principles.  
In the same spirit, the radiation line width
of a Josephson junction was shown to originate from the voltage 
fluctuations of the external circuit \cite{larkin}.
A fundamental question here is to derive
this line width and therefore to see how dissipation affects the measurement 
of the higher current moments.            

\begin{figure}
\begin{center}
	\includegraphics[width=7.5cm]{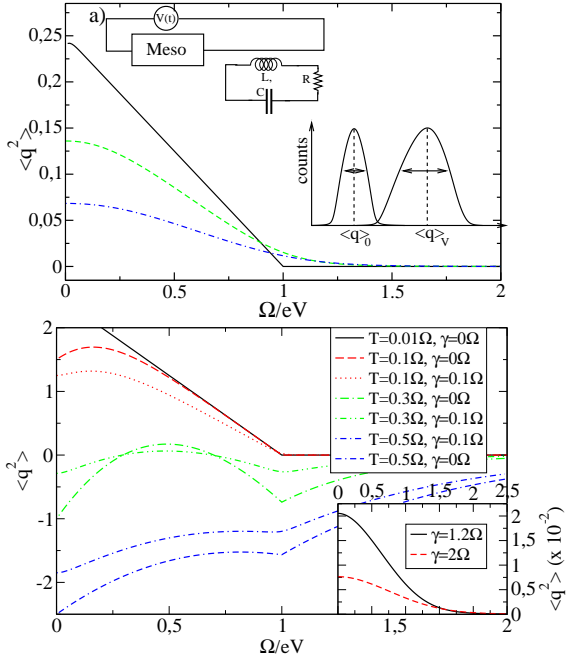}
	\end{center}
\caption{(color online) a) (top) Mesoscopic device coupled to a dissipative LC circuit.
(bottom) typical histograms of the charge used to identify the noise and the third moment, at zero and finite voltage.
Measured noise (damping parameter $\gamma$) at temperature $T=0.01\Omega$.
No damping (full line), $\gamma/2=0.4\Omega$ (dashed line), $\gamma/2=0.6\Omega$
(dashed-dotted line). b) $T>\gamma/2$ (underdamped case). Inset:  $T=0.01\Omega$ 
(over-damped regime).}\label{fig1}
\end{figure}

The setup is depicted in the upper part of Fig. \ref{fig1}a: a lead 
from the mesoscopic device is inductively coupled to a resonant circuit
(capacitance $C$, inductance $L$, and dissipative component $R$).   
Repeated time measurements are operated on the charge $q$, 
which yield an histogram which is qualitatively depicted in Fig. \ref{fig1}a:  
a reference histogram is made for zero voltage (left), yielding 
the zero bias peak position, its width, its skewness,... In the presence of bias, 
this histogram is shifted (right), and it acquires a new width. Information 
about all current moments at high frequencies is coded in such histograms.    
The basic Hamiltonian which describes the oscillator (the LC circuit) reads:
$H_{osc}=H_0+V$
where
$H_0=H_{free}+H_{bath}$ is the Hamiltonian of the uncoupled system.
We use a path integral
formulation to describe the evolution of the oscillator in the presence 
of coupling to the bath and to the mesoscopic device.  
In the absence of coupling the Keldysh action describing the charge of the LC 
circuit reads: 
\begin{eqnarray}
S_{osc}[q]=\frac{1}{2}\int_{-\infty}^{+\infty}dt dt'{\bf q}^T(t) G_0^{-1}(t-t')\sigma_z {\bf q}(t')~,
\end{eqnarray}
with the Green's function
$G_0^{-1}(t)=M[( i\partial_t)^2-\Omega^2]$, $\Omega=(LC)^{-1/2}$ is the resonant 
frequency of the circuit. ${\bf q}^T=(q^+,q^-)$ is a two component vector which contains 
the oscillator coordinate on the forward/backward contour, $\sigma_z$ is a Pauli matrix
in Keldysh space.
The action describing the free LC circuit is that of an harmonic oscillator. 
Dissipative effects are treated within the Caldeira-Leggett model \cite{caldeira}: 
$q$ is coupled to an oscillator bath, whose 
coordinate $x_n$ 
has a Green's function $D^{-1}_n(t)=M_n[( i\partial_t)^2-\Omega_n^2]$
(same as the undamped circuit). 
The coupling between  $q$ and $x_n$ is chosen to be linear, 
$V=q\sum_n{\lambda_nx_n}$.
The partition function of the oscillator plus bath
$Z=\int\mathcal{D}q\mathcal{D}xe^{ i S[q,x]}$ has an action:
\begin{equation}
   \begin{split}
    S[{\bf q},{\bf x}]=&S_{osc}[{\bf q}]+
\frac{1}{2}\sum_n{\bf x}^T_n\circ D_n^{-1}\circ \sigma_z {\bf x}_n\\
    &~~-
    {\bf q}^T\circ\sigma_z
\sum_n\lambda_n {\bf x}_n~,
   \end{split}
   \end{equation}
where the symbol $\circ$ stands for convolution in time.
The environment degrees of freedom being quadratic, they can be integrated out in a standard manner \cite{grabert_report}.
The Green's function of the LC circuit becomes dressed by its electronic environment,
$G^{-1}(t-t')=G_0^{-1}(t-t')-\Sigma(t-t')$, with a 
self energy $\Sigma(t-t')=\sigma_z\sum_n\lambda_n^2D_n\sigma_z$. 
    
Next, we introduce the inductive
coupling between the mesoscopic circuit
and the LC circuit
$V_{int}=\alpha q \dot{I}$, where $\dot{I}$ is the time derivative 
of the current operator \cite{lesovik_loosen}. This interaction is interpreted 
here as an external potential acting on the oscillator circuit. 
Because we are interested in calculating correlation functions
of the LC circuit coordinate, we introduce a two-component auxiliary field
$\eta$ (upper/lower contour) which allows to write the partition function:
\begin{eqnarray}
\mathcal{Z}_\eta[I]=\int \mathcal{D}{\bf q}^T \exp i \Big[\frac{1}{2}{\bf q}^T\circ G^{-1} \circ {\bf q} -{\bf q}^T\sigma_z\circ(\alpha \dot{\bf I}+{\bf \eta})\Big]~.
\end{eqnarray}

The effective action is then quadratic in the oscillator coordinate, so that one can integrate out 
${\bf q}$, and the effective action becomes (restoring integrals): 
 \begin{eqnarray}
S_{eff}&=&-\frac{ i}{2}\int dt \int dt' ({\bf \eta}(t)+\alpha\dot{\bf I}(t))^T\sigma_z
\check{G}(t-t')\nonumber\\
&&~~~~~~\times \sigma_z ({\bf \eta}(t')+\alpha\dot{\bf I}(t'))\Big]~.
 \label{action_integrated}
 \end{eqnarray}
The action of Eq. (\ref{action_integrated}) is then used to compute 
the relevant averages by taking derivatives over the auxiliary field:
%
\begin{eqnarray}
&&\langle q(t) \rangle
=\frac{\alpha}{2}\sum_{s}\int d\tau\sigma_z^{s,s}(G^{+s}(t,\tau)+G^{-s}(t,\tau))
\nonumber\\
&&~~~~~~~~\times
\langle \dot{I}(\tau^{s})Z[I]\rangle/\langle Z[I]
\rangle~,
\label{exact_position}
\\
&&\delta \langle q(t)q(0) \rangle
= \alpha^2 \sum_{s_1,s_2} \sigma_z^{s_1s_1} \sigma_z^{s_2s_2} 
Z[I]^{-1}
\int d \tau_1 d \tau_2 \nonumber\\
&&~~~~\times\langle\dot{I} (\tau_1^{s_1}) \dot{I} (\tau_2^{s_2})Z[I]\rangle G^{s_1+} (\tau_1,0) G^{s_2-} (\tau_2,t)~,
\label{exact_fluctu}
\end{eqnarray}
with $Z[I]=\mathcal{Z}_{\eta=0}[I]$, and $\langle...\rangle$ denotes a 
non-equilibrium average over the mesoscopic system.
In the above, we ignore contributions which originate from the zero point 
fluctuations of the LC circuit plus bath, as these are subtracted in the excess  
noise and third moment measurement which is implied in Fig \ref{fig1}.  
At this stage no approximation has been made on the magnitude of the 
inductive coupling. An expansion of the partition function in powers of 
$\alpha$ yields contributions for these averages
which contain all high-order correlators of the current derivative moments. 
Such moments are translated into ``regular'' 
current correlators, using Fourier 
transforms.  We start with noise,
introducing the combination: 
$K^\pm(t)=\theta(t)(K^>(t)\pm K^<(t))$, where $K^>$ and $K^<$ are the off diagonal
elements of $\langle \dot{I}(t^s) \dot{I}(t'^{s'}) \rangle$
in Eq. (\ref{exact_fluctu}). Going to the rotated Keldysh basis 
allows to rewrite the charge fluctuations at equal time as
(the time dependance drops out):
\begin{eqnarray}
\delta\langle q^2 \rangle &=& \alpha^2 \int \frac{d\omega}{2\pi} G^R(\omega) \{G^K(\omega)K^-(\omega)\nonumber\\
&&-(G^R(\omega)-G^A(\omega))K^+(\omega)\}~,
\label{general_second}
\end{eqnarray}
with the three non-vanishing Green's oscillator function: 
$G^{R/A}(\omega)
=[M(\omega^2-\Omega^2)\pm i \textrm{sgn}(\omega)J(|\omega|)]^{-1}$
and
$G^K=(2N(\omega)+1)(G^R(\omega)-G^A(\omega))
$
($N(\omega)$ is the Bose-Einstein distribution),
were the spectral function of the bath
$J(\omega)=\pi\sum_n\lambda_n^2/(2M_n\Omega_n)\delta(\omega-\Omega_n)$  
gives rise to a finite line width for the LC circuit Green's function. 

Next, we relate the time derivative correlators to 
the current correlators:
$K^<(\omega)=\omega^2S_+(\omega)$ and 
$K^>(\omega)=\omega^2S_-(\omega)$, 
with $S_+(\omega)= \int dt \langle I(0) I(t)\rangle e^{i\omega t}$ and
$S_-(\omega)=S_+(-\omega)$, which correspond to the response function for 
emission/absorption of radiation from/to the mesoscopic circuit 
\cite{lesovik_loosen,deblock_science}. 
With these definitions, the final result for the measured excess noise reads: 
\begin{eqnarray}
\delta\langle q^2 \rangle&=& 
  2\alpha^2 \int_{0}^{\infty} \frac{d\omega}{2\pi}
 \omega^2
 [\chi''(\omega)]^2
\nonumber\\
&&~~\times 
  \big(S_+(\omega)+N(\omega)(S_+(\omega)-S_-(\omega))\big)~,
\label{ohmic_second}  
\end{eqnarray} 
where $\chi''(\omega)= J(|\omega|)/[M^2(\omega^2-\Omega^2)^2 + J^2(|\omega|)$
is the susceptibility of Ref. \cite{caldeira}, here generalized to arbitrary
$J(|\omega|)$. 
Eq. (\ref{ohmic_second}) indicates that for a small line width, 
the integrand can be computed at the resonant frequency $\Omega$, and the 
measured noise is proportional to $S_+(\Omega)+N(\Omega)(S_+(\Omega)-S_-(\Omega))$,
with a prefactor which diverges when the circuit is uncoupled to its environment \cite{lesovik_loosen}. 
Eq. (\ref{ohmic_second})
constitutes a mesoscopic analog of the radiation line width
calculation of \cite{larkin}: a dissipative LC circuit cannot yield any divergences
in the measured noise. Dissipation is essential in the measurement 
process.

Next, we turn to the measurement of the third moment. Performing a perturbative expansion in $\alpha$
of the average charge in Eq (\ref{exact_position}), only odd current (derivative)
correlators can be generated in this series. The first term is proportional 
to $\langle \dot{I} \rangle$: it vanishes in a stationary situation (DC bias 
on the mesoscopic device). The next non-vanishing term is directly related to the third moment 
at finite frequencies:   
$
L^{s_1,s_2,s_3}(t_1,t_2,t_3)=\langle T_K\{\dot{I}(t_1^{s_1} \dot{I}(t_2^{s_2})\dot{I}(t_3^{s_3})\}\rangle
$.
The average charge is expressed in terms 
of the Green's functions of the LC circuit
plus bath and the current correlators.
\begin{eqnarray}
&&\langle q \rangle_{(3)}=-\frac{ i}{2}
\alpha^3\int d\tau\theta(t-\tau)(G^>(t,\tau)-G^<(t,\tau))\nonumber\\
&&\times\int
dt_1 dt_2 \sum_{s_1,s_2}\sigma_z^{s_1s_1}\sigma_z^{s_2s_2}G^{s_1s_2}(t_1-t_2)L^{+s_1s_2}(\tau,t_1,t_2)~.
\end{eqnarray}
It turns out that the average charge can be expressed solely in terms of a special combination of current
derivative correlators:
\begin{eqnarray}
R^\pm(\tau,t_1,t_2)&=&\theta(\tau-t_1)\theta(t_1-t_2)L^\pm(\tau,t_1,t_2)~,
\end{eqnarray}
with
\begin{eqnarray}
L^\mp(\tau,t_1,t_2)&=&\LRangle{\Big[[\dI{\tau},\dI{t_1}]_-,\dI{t_2}\Big]_\mp}~.
\label{commutator}
\end{eqnarray}
That is, the mesoscopic circuit correlators appear only in the form of interlocked 
commutators ($-$)/anti-commutators ($+$). 
This is an important aspect of this scheme, because the commutator which is common for both
correlators of Eq. (\ref{commutator}) implies that our scheme is only effective
when the transport is fully coherent, i.e. when the rate of escape for electrons from
the mesoscopic device to the leads is large compared to the temperature.  
Such correlators vanish in the case of incoherent Coulomb blockade transport.
Exploiting time translational invariance, the final result for the measured 
third moment then reads:     
\begin{eqnarray}
\lrangle{q}_{(3)}&=&-i\alpha^3  \int_{-\infty}^{\infty} \frac{ d\omega }{2\pi}
 G^R(0)\Big[
G^K(\omega)R^-(0,\omega)\nonumber\\
&&~~~~~~-(G^R(\omega)-G^A(\omega))R^+(0,\omega)\Big]~,
\label{general_third}
\end{eqnarray}
Note the similarity between 
this expression and the one obtained in Eq. (\ref{general_second}) for the measured noise.
$R^+$ is weighted by same the spectral density of states of the LC oscillator plus bath 
$G^R-G^A$, as for the factor $K^+$ in Eq. (\ref{general_second}). The factor $R^-$, 
on the other hand, does not contain the same frequencies as that of $K^-$ (see below). 
This result also shows which third moment correlator (which frequencies) the dissipative LC 
circuit is capable of measuring.      
Using the expressions of the oscillator circuit Green's function:
 \begin{eqnarray}
&&\lrangle{q}_{(3)}
=  
\frac{2\alpha^3}{M\Omega^2}
\int_{0}^{\infty}  d\omega \chi''(\omega)
 \nonumber\\
&&~~~~\times  
\textrm{Re}\Big\{ \Big[
(2N(\omega)+1)R^-(0,\omega)-R^+(0,\omega)\Big]\Big\}~,
\label{ohmic_third}
 \end{eqnarray}

There is a fundamental difference between the two responses
of Eqs. (\ref{ohmic_second}) and (\ref{ohmic_third}). 
$\chi''(\omega)$ appears as a square in 
the measured noise (\ref{ohmic_second}), while it 
does not in the third moment (\ref{ohmic_third}). 
We next specify to strict ohmic or Markovian damping
$J(\omega)= M \gamma \omega$, a memoryless bath being consistent with the
adiabatic switching assumption.
The limit of zero ohmic dissipation $\gamma\to 0$ leads to a 
divergence in the measured noise because it is proportional to the square
of $\chi''(\omega)$: there are ``large'' charge 
fluctuations on the capacitor plates because it is pumping energy from the 
mesoscopic device. A finite damping is needed in order for the 
integral in Eq. (\ref{ohmic_second}) to converge: it explains the breakdown of 
adiabaticity\cite{lesovik_loosen}. The measured 
third moment is not singular when $\gamma\to 0$. 

\begin{figure}
\begin{center}
\includegraphics[width=7.5cm]{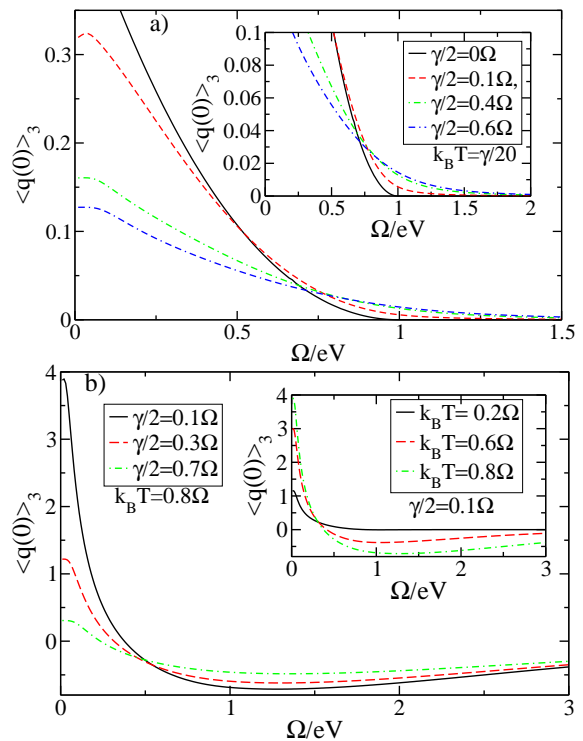}
\end{center}
\caption{(color online) Measured third moment. a) $T\ll \gamma/2$ and $\gamma/2< \Omega$
for the displayed values of the damping parameter. Inset: zoom near $\Omega=eV$.
b) case $\gamma/2<T$, at constant temperature and different value of $\gamma$. Inset:
$\gamma$ is fixed and $T$ is varied. }\label{fig2}
\end{figure} 

Results are applied to a point contact. 
The excess noise is known to have a singular derivative at $\omega=eV$.    
$\langle q^2\rangle$ and $\langle q\rangle$ are plotted as a function of $\Omega/eV$, $\gamma/\Omega$, 
and $T/\Omega$, with $T$ the temperature of the LC circuit, which is assumed to be
small compared to $eV$ (shot noise dominated regime) \cite{footnote}.
In the under-damped case the susceptibility $\chi''$ is a superposition of
Lorentzian peaks at $\pm \Omega$ and width $\gamma$. Thus, if $\gamma \ll \Omega$
we expect qualitative behavior similar to that of the undamped case,
which is indeed what happens, with the important result that the divergency
is removed. 
Fig. \ref{fig1}a shows that at small temperature, the effect of damping 
is to wash out the singularity, and the measured noise flattens out. A curve with no damping
is shown for comparison, after rescaling (it is infinite at $\omega=0$ for $\gamma\to 0$).  
The inset of Fig. \ref{fig1}b also applies to $T<\gamma$, but deals with the 
over damped regime: there is no reminiscence of the linear behavior found in the absence 
of damping because the two peaks of $\chi''(\omega)$ cannot be resolved, even at low temperatures. 
Fig. \ref{fig1}b shows the effect of the temperature on the noise both without and with dissipation, 
in the under damped regime. The measured noise can become negative at higher temperature 
because $S_+-S_-<0$, and because of the large population of LC oscillator states. 
Because we are considering excess effects (difference between the charge fluctuations 
with and without the applied bias) there is no controversy here. 
As in Fig. \ref{fig1}a, the cusps (or singularities), which survive for the undamped case 
even at these temperatures, are strongly attenuated due to damping.
An important feature is that the measuring temperature $T$ enters our results
exactly as in the undamped case, because the response function $\chi''(\omega)$
is temperature-independent ($\chi''(\omega)$ is related to the symmetrized
correlation function of the damped HO via the fluctuaction-dissipation
theorem).

We turn now to the measured third moment (Fig. \ref{fig2}). For $\gamma\to 0$,     
it does not have a singularity at $\omega=eV$, but it vanishes beyond 
this point, and has a linear behavior (not shown) close to $\omega=0$. 
For the under damped case $\gamma/2<\Omega$, the main effect is to reduce the amplitude
of the measured third moment, and to wash out its vanishing at $\Omega=eV$ (see inset). 
Furthermore, one notices that the third moment saturates near $\omega=0$, and acquires 
a maximum in this region. The effect of temperature is displayed in Fig. \ref{fig2}:
the structure at $\Omega=eV$ disappears, and the width of the maximum at $\omega=0$ is 
reduced.  Similarly to the measured noise, the third moment can become negative either when 
the damping is increased (Fig. \ref{fig2}b), or when the temperature is increased
(inset). 

The above measurement setup and  coupling
conditions are easily achievable by on-chip inductive coupling to
a SQUID circuit behaving as a harmonic oscillator. Recently reported
quality factors of $\approx 100 - 150$, with an oscillator resonance
of $\approx 3 $GHz, and operating temperature $T \approx 25 mk$
\cite{chiorescu_johansson} correspond to the under damped regime
discussed. A detailed investigation of the experimental setup
will be reported elsewhere.

In summary, dissipation was included in the measurement 
of the higher current moments in the coherent regime; it is crucial to get a 
finite result for the noise. Third moment correlators
have been identified with this scheme. 
The measurement of higher current moments, using the skewness and the sharpness
of the charge histogram consitutes an extension. 

T.M. and A.C. acknowledge support an ANR grant from 
the French ministry of research. 
E.P. thanks CPT for its hospitality. Discussions with G. Falci 
are gratefully acknowledged.


\begin{thebibliography}{99}

\bibitem{reulet}  B. Reulet, J. Senzier, and D. E. Prober
Phys. Rev. Lett. {\bf 91}, 196601 (2003) 

\bibitem{reznikov}  Yu. Bomze, G. Gershon, D. Shovkun, L. S. Levitov, and M. Reznikov
Phys. Rev. Lett. {\bf 95}, 176601 (2005) 

\bibitem{fujisawa} T. Fujisawa, {\it et al.},
Science {\bf 312}, 1634 (5780);
S. Gustavsson {\it et al.},
Phys. Rev. Lett. {\bf 96}, 076605 (2006).

  
S. Gustavsson, R. Leturcq, T. Ihn, K. Ensslin, M. Reinwald, W. Wegscheider, 
cond-mat/0607192.


\bibitem{onac} 
E. Onac, F. Balestro, L. H. W. van Beveren, U. Hartmann, Y. V. Nazarov, and L. P. Kouwenhoven
Phys. Rev. Lett. {\bf 96}, 176601 (2006);
E. Onac, F. Balestro, B. Trauzettel, C. F. J. Lodewijk, and L. P. Kouwenhoven
ibid. {\bf 96}, 026803 (2006) 
\bibitem{lesovik_loosen} G. B. Lesovik and R. Loosen, Pis'ma Zh. \'Eksp. 
Teor. Fiz. {\bf 65}, 280 (1997) [JETP Lett. {\bf 65}, 295, (1997)];
U. Gavish, I. Imry, and Y. Levinson,
Phys. Rev. B {\bf 62}, 10637 (2000). 

\bibitem{aguado_kouwenhoven}R. Aguado and L. P. Kouwenhoven, Phys. Rev. Lett. 
{\bf 84}, 1986 (2000).

\bibitem{deblock_science} R. Deblock, E. Onac, L. Gurevich, and L. P. Kouvenhoven,
Science {\bf 301}, 203 (2003);
%
P.-M. Billangeon, F.Pierre, H. Bouchiat, and R. Deblock, Phys. Rev. Lett. 
{\bf 96}, 136804 (2006).

\bibitem{trauzettel_lebedev}  B. Trauzettel, I. Safi, F. Dolcini, and H. Grabert
Phys. Rev. Lett. 92, 226405 (2004); A. V. Lebedev, A. Cr\'epieux, and T. Martin,
Phys. Rev. B {\bf 71}, 075416 (2005).


\bibitem{beenaker} C. W. J. Beenakker, M. Kindermann, and Yu. V. Nazarov, Phys.
Rev. Lett. {\bf 90}, 176802 (2003). 

\bibitem{other}  J. Tobiska and Yu. V. Nazarov
Phys. Rev. Lett. {\bf 93}, 106801 (2004); J. P. Pekola, {\it ibid.}, {\bf 93}, 206601 (2006);
T. T. Heikkil\"a, P. Virtanen, G. Johansson, and F. K. Wilhelm
{\it ibid.}, {\bf 93}, 247005 (2004); 
T. Ojanen and T. T. Heikkil\"a
Phys. Rev. B {\bf 73}, 020501 (2006); 
V. Brosco, R. Fazio, F. W. J. Hekking, and J. P. Pekola
{\it ibid.}, {\bf 74}, 024524 (2006).

\bibitem{gabelli} J. Gabelli, L.-H. Reydellet, G. Feve, J.-M. Berroir, B. Pla\c{c}ais, P. Roche, and D. C. Glattli,
Phys. Rev. Lett. {\bf 93}, 056801 (2004). 

\bibitem{larkin} A.I. Larkin and Yu. N. Ovchinikov, Zh. Eksp. Teor. Fiz. {\bf 53}, 2159 (1967) 
[Sov. Phys. JETP {\bf 26}, 1219 (1968)].

\bibitem{caldeira} A. O. Caldeira and A. J. Leggett, Physica {121}A, 587 (1983). 

\bibitem{grabert_report}  H. Grabert, P. Schramm, and G.-L. Ingold
Phys. Rep. {\bf 168}, 115 (1988).

\bibitem{footnote} When the current is amplified before being coupled to the LC circuit, this assumption 
can be relaxed, but additional filtering due to the amplifiers occurs.



\bibitem{chiorescu_johansson}
I. Chiorescu et.al., Nature, 431, 159, (2004);
J. Johansson {\it et al.}, Phys. Rev. Lett., 96, 127006 (2006).




 





 

























\end{thebibliography}
\end{document}